# OPTIMAL SYNTHESIS OF MULTIPLE ALGORITHMS

KERRY M. SOILEAU

ksoileau@yahoo.com

**ABSTRACT**

In this paper we give a definition of "algorithm," "finite algorithm," "equivalent algorithms," and what it means for a single algorithm to dominate a set of algorithms. We define a derived algorithm which may have a smaller mean execution time than any of its component algorithms. We give an explicit expression for the mean execution time (when it exists) of the derived algorithm. We give several illustrative examples of derived algorithms with two component algorithms. We include mean execution time solutions for two-algorithm processors whose joint density of execution times are of several general forms. For the case in which the joint density for a two-algorithm processor is a step function, we give a maximum-likelihood estimation scheme with which to analyze empirical processing time data.

# 1 INTRODUCTION

It can categorically be said that no algorithm is unique. By this we mean that for a given task, invariably more than one algorithm exists which will accomplish that task. One strategy is to select one algorithm deemed generally superior to the rest, and to use that algorithm exclusively. This paper examines an alternative strategy. We ask, given two or more equivalent algorithms, is it ever possible to create a new derived algorithm whose mean execution time is less than that of all of the original algorithms? If so, how can such an algorithm be derived?

First we define clearly what we mean by the term "algorithm:"

<u>Algorithm</u>: An algorithm $\alpha$ is a pair $(\rho_\alpha, \pi_\alpha)$, where $\rho_\alpha : \Omega \to \Gamma$ is a Turing-computable mapping of a countable set $\Omega$ (tasks) into a countable set $\Gamma$ (outputs), and $\pi_\alpha : \Omega \to \mathbb{R}$ is a mapping of $\Omega$ into the positive real numbers. The function $\rho_\alpha$ specifies the algorithm's output $\rho_\alpha(\omega)$ when presented with the task $\omega \in \Omega$. The function $\pi_\alpha$ specifies the execution time $\pi_\alpha(\omega)$ required to compute the output $\rho_\alpha(\omega)$. Note that under this definition, given a task $\omega \in \Omega$, an algorithm will always produce a definite output, namely $\rho_\alpha(\omega)$, and will always produce this output after a definite amount of time has passed, namely $\pi_\alpha(\omega)$. We do not address procedures which are nondeterministic or whose execution time is unpredictable.

<u>Definition</u>: We say that an algorithm $\alpha = (\rho_\alpha, \pi_\alpha)$ is <u>finite</u> if and only if $0 < \pi_\alpha(\omega) < \infty$ for every $\omega \in \Omega$. Note that "$\alpha = (\rho_\alpha, \pi_\alpha)$ is finite" does not imply "$\pi_\alpha$ is bounded." For example, Quicksort and Bubblesort are finite.

Definition: We say that two algorithms $\alpha = (\rho_\alpha, \pi_\alpha)$ and $\beta = (\rho_\beta, \pi_\beta)$ are underline{equivalent} if and only if $\text{Dom } \rho_\alpha = \text{Dom } \rho_\beta$ and $\rho_\alpha(\omega) = \rho_\beta(\omega)$ for every $\omega \in \Omega$. Notice that equivalent algorithms may require different times to process a given task. For example, Quicksort and Bubblesort are equivalent.

Definition: Let $\{\alpha_1, \alpha_2, \cdots, \alpha_N\}$ be a set of equivalent algorithms. We say that $\alpha_n$ underline{dominates} $\{\alpha_1, \alpha_2, \cdots, \alpha_N\}$ if and only if for every $\omega \in \Omega$, $\pi_{\alpha_n}(\omega) \leq \pi_{\alpha_i}(\omega)$ for every $i \in \{1, 2, \cdots, N\}$.

Now suppose we are given a set of finite equivalent algorithms $\{\alpha_1, \alpha_2, \cdots, \alpha_N\}$ such that no $\alpha_n$ dominates $\{\alpha_1, \alpha_2, \cdots, \alpha_N\}$. Suppose further that there exists a probability space $(\Omega, \Im, P)$ over $\Omega$ such that $\pi_{\alpha_1}, \pi_{\alpha_2}, \cdots, \pi_{\alpha_N}$ are random variables. Let $f_{\pi_{\alpha_1}, \pi_{\alpha_2}, \cdots, \pi_{\alpha_N}} : \mathbb{R}^N \to \mathbb{R}$ be the joint density of the random variables $\pi_{\alpha_1}, \pi_{\alpha_2}, \cdots, \pi_{\alpha_N}$.

Definition of Derived Algorithm: From a set of finite equivalent algorithms $\{\alpha_1, \alpha_2, \cdots, \alpha_N\}$, and a given point $(\tau_1, \tau_2, \cdots, \tau_{N-1}) \in [0, \infty)^{N-1}$, the function $\left[\alpha_1 \mid_{\tau_1} \alpha_2 \mid_{\tau_2} \cdots \alpha_{N-1} \mid_{\tau_{N-1}} \alpha_N \right] : \Omega \to \Gamma$ is defined as follows. For each $(\tau_1, \tau_2, \cdots, \tau_{N-1}) \in [0, \infty)^{N-1}$, we define the random variable

$$T_{\pi_{\alpha_1}, \pi_{\alpha_2}, \cdots, \pi_{\alpha_N}}(\tau_1, \tau_2, \cdots, \tau_{N-1}) : \Omega \to \mathbb{R}$$

as follows:

$$T_{\pi_{\alpha_1},\pi_{\alpha_2},\cdots,\pi_{\alpha_N}}(\tau_1,\tau_2,\cdots,\tau_{N-1})(\omega)$$

$$=\begin{cases} \pi_{\alpha_1}(\omega) & \omega \in \Omega \sim S_1 \\ \tau_1 + \pi_{\alpha_2}(\omega) & \omega \in S_1 \sim S_2 \\ \tau_1 + \tau_2 + \pi_{\alpha_3}(\omega) & \omega \in S_2 \sim S_3 \\ \vdots & \vdots \\ \tau_1 + \tau_2 + \cdots + \tau_{N-2} + \pi_{\alpha_{N-1}}(\omega) & \omega \in S_{N-2} \sim S_{N-1} \\ \tau_1 + \tau_2 + \cdots + \tau_{N-1} + \pi_{\alpha_N}(\omega) & \omega \in S_{N-1} \end{cases} \quad (1)$$

and

$$\left[\alpha_1 \mid_{\tau_1} \alpha_2 \mid_{\tau_2} \cdots \alpha_{N-1} \mid_{\tau_{N-1}} \alpha_N\right](\omega) = \begin{cases} \rho_{\alpha_1}(\omega) & \omega \in \Omega \sim S_1 \\ \rho_{\alpha_2}(\omega) & \omega \in S_1 \sim S_2 \\ \rho_{\alpha_3}(\omega) & \omega \in S_2 \sim S_3 \\ \vdots & \vdots \\ \rho_{\alpha_{N-1}}(\omega) & \omega \in S_{N-2} \sim S_{N-1} \\ \rho_{\alpha_{N-1}}(\omega) & \omega \in S_{N-1} \end{cases} \quad (2)$$

where $S_n = \{\omega \in \Omega; \tau_1 < \pi_{\alpha_1}(\omega), \tau_2 < \pi_{\alpha_2}(\omega), \cdots, \tau_n < \pi_{\alpha_n}(\omega)\}$ for $1 \le n \le N-1$. Each $S_n$ is the event consisting of the points $\omega \in \Omega$ on which none of the algorithms $\alpha_1, \alpha_2, \cdots, \alpha_n$ completes processing within each algorithm's permitted run time limit. The <u>derived algorithm</u> is then defined to be the pair

$$\left(\left[\alpha_1 \mid_{\tau_1} \alpha_2 \mid_{\tau_2} \cdots \alpha_{N-1} \mid_{\tau_{N-1}} \alpha_N\right], T_{\pi_{\alpha_1},\pi_{\alpha_2},\cdots,\pi_{\alpha_N}}(\tau_1,\tau_2,\cdots,\tau_{N-1})\right).$$

$T_{\pi_{\alpha_1},\pi_{\alpha_2},\cdots,\pi_{\alpha_N}}(\tau_1,\tau_2,\cdots,\tau_{N-1})(\omega)$ represents the time taken for the derived algorithm to execute when presented with the task $\omega$, and $\left[\alpha_1 \mid_{\tau_1} \alpha_2 \mid_{\tau_2} \cdots \alpha_{N-1} \mid_{\tau_{N-1}} \alpha_N\right]$ represents the derived algorithm's output when presented with the task $\omega$.

We may envision an implementation of this algorithm as follows. When presented with a task $\omega \in \Omega$, a timer is started, and $\alpha_1$ is applied. If $\alpha_1$ has not completed by time $\tau_1$, $\alpha_1$ is abandoned and $\alpha_2$ is applied. If $\alpha_2$ has not completed by time $\tau_1 + \tau_2$, $\alpha_2$ is abandoned and $\alpha_3$ is applied, and so on. If $\alpha_{N-1}$ has not completed by time $\tau_1 + \tau_2 + \cdots + \tau_{N-1}$, $\alpha_{N-1}$ is abandoned and $\alpha_N$ is applied and (unlike the other algorithms) is allowed to run without time limit. $\rho_{\alpha_i}(\omega)$ is returned as output, where $\alpha_i$ is the algorithm which completed execution on the task $\omega \in \Omega$.

The expected value (if it exists) of the random variable $T_{\pi_{\alpha_1}, \pi_{\alpha_2}, \cdots, \pi_{\alpha_N}}(\tau_1, \tau_2, \cdots, \tau_{N-1})$ is given by the following

**Theorem 1**:

$$ET_{\pi_{\alpha_1}, \pi_{\alpha_2}, \cdots, \pi_{\alpha_N}}(\tau_1, \tau_2, \cdots, \tau_{N-1}) = \left(E\pi_{\alpha_1} | \Omega \sim S_1\right) P(\Omega \sim S_1)$$
$$+ \sum_{n=2}^{N-1} \left(E\pi_{\alpha_n} | S_{n-1} \sim S_n\right) P(S_{n-1} \sim S_n) + \left(E\pi_{\alpha_N} | S_{N-1}\right) P(S_{N-1}) \quad (3)$$
$$+ \sum_{n=1}^{N-1} \tau_n P(S_n)$$

Proof: Recall that

$$T_{\pi_{\alpha_1}, \pi_{\alpha_2}, \cdots, \pi_{\alpha_N}}(\tau_1, \tau_2, \cdots, \tau_{N-1})(\omega)$$
$$= \begin{cases} \pi_{\alpha_1}(\omega) & \omega \in \Omega \sim S_1 \\ \tau_1 + \pi_{\alpha_2}(\omega) & \omega \in S_1 \sim S_2 \\ \tau_1 + \tau_2 + \pi_{\alpha_3}(\omega) & \omega \in S_2 \sim S_3 \\ \vdots & \vdots \\ \tau_1 + \tau_2 + \cdots + \tau_{N-2} + \pi_{\alpha_{N-1}}(\omega) & \omega \in S_{N-2} \sim S_{N-1} \\ \tau_1 + \tau_2 + \cdots + \tau_{N-1} + \pi_{\alpha_N}(\omega) & \omega \in S_{N-1} \end{cases}$$

It follows immediately that

$$ET_{\pi_{\alpha_1},\pi_{\alpha_2},\cdots,\pi_{\alpha_N}}(\tau_1,\tau_2,\cdots,\tau_{N-1})$$
$$=(E\pi_{\alpha_1}|\Omega \sim S_1)P(\Omega \sim S_1)$$
$$+E(\tau_1+\pi_{\alpha_2}|S_1 \sim S_2)P(S_1 \sim S_2)$$
$$+E(\tau_1+\tau_2+\pi_{\alpha_3}|S_2 \sim S_3)P(S_2 \sim S_3) \quad (4)$$
$$+$$
$$\cdots$$
$$+E(\tau_1+\tau_2+\cdots+\tau_{N-2}+\pi_{\alpha_{N-1}}|S_{N-2} \sim S_{N-1})P(S_{N-2} \sim S_{N-1})$$
$$+E(\tau_1+\tau_2+\cdots+\tau_{N-1}+\pi_{\alpha_N}|S_{N-1})P(S_{N-1})$$

This may be written as

$$ET_{\pi_{\alpha_1},\pi_{\alpha_2},\cdots,\pi_{\alpha_N}}(\tau_1,\tau_2,\cdots,\tau_{N-1}) = (E\pi_{\alpha_1}|\Omega \sim S_1)P(\Omega \sim S_1)+\tau_1 P(S_1 \sim S_2)$$
$$+(E\pi_{\alpha_2}|S_1 \sim S_2)P(S_1 \sim S_2)+(\tau_1+\tau_2)P(S_2 \sim S_3)$$
$$+(E\pi_{\alpha_3}|S_2 \sim S_3)P(S_2 \sim S_3)$$
$$+\cdots+(\tau_1+\tau_2+\cdots+\tau_{N-2})P(S_{N-2} \sim S_{N-1}) \quad (5)$$
$$+(E\pi_{\alpha_{N-1}}|S_{N-2} \sim S_{N-1})P(S_{N-2} \sim S_{N-1})$$
$$+(\tau_1+\tau_2+\cdots+\tau_{N-1})P(S_{N-1})+(E\pi_{\alpha_N}|S_{N-1})P(S_{N-1})$$

Telescoping sums yield

$$ET_{\pi_{\alpha_1},\pi_{\alpha_2},\cdots,\pi_{\alpha_N}}(\tau_1,\tau_2,\cdots,\tau_{N-1}) = (E\pi_{\alpha_1}|\Omega \sim S_1)P(\Omega \sim S_1)$$
$$+(E\pi_{\alpha_2}|S_1 \sim S_2)P(S_1 \sim S_2)$$
$$+(E\pi_{\alpha_3}|S_2 \sim S_3)P(S_2 \sim S_3)+\cdots$$
$$+(E\pi_{\alpha_{N-1}}|S_{N-2} \sim S_{N-1})P(S_{N-2} \sim S_{N-1}) \quad (6)$$
$$+(E\pi_{\alpha_N}|S_{N-1})P(S_{N-1})$$
$$+\tau_1(P(S_1 \sim S_2)+P(S_2 \sim S_3)+\cdots+P(S_{N-2} \sim S_{N-1})+P(S_{N-1}))$$
$$+\tau_2(P(S_2 \sim S_3)+\cdots+P(S_{N-2} \sim S_{N-1})+P(S_{N-1}))+\cdots+\tau_{N-1}(P(S_{N-1}))$$

Next,

$$\begin{aligned}
ET_{\pi_{\alpha_1},\pi_{\alpha_2},\cdots,\pi_{\alpha_N}}(\tau_1,\tau_2,\cdots,\tau_{N-1}) &= \left(E\pi_{\alpha_1}\big|\Omega \sim S_1\right)P(\Omega \sim S_1) \\
&+ \left(E\pi_{\alpha_2}\big|S_1 \sim S_2\right)P(S_1 \sim S_2) \\
&+ \left(E\pi_{\alpha_3}\big|S_2 \sim S_3\right)P(S_2 \sim S_3) + \cdots + \left(E\pi_{\alpha_{N-1}}\big|S_{N-2} \sim S_{N-1}\right)P(S_{N-2} \sim S_{N-1}) \\
&+ \left(E\pi_{\alpha_N}\big|S_{N-1}\right)P(S_{N-1}) + \tau_1 P(S_1) + \tau_2 P(S_2) + \cdots + \tau_{N-1} P(S_{N-1})
\end{aligned} \quad (7)$$

whence

$$\begin{aligned}
ET_{\pi_{\alpha_1},\pi_{\alpha_2},\cdots,\pi_{\alpha_N}}(\tau_1,\tau_2,\cdots,\tau_{N-1}) &= \left(E\pi_{\alpha_1}\big|\Omega \sim S_1\right)P(\Omega \sim S_1) \\
&+ \sum_{n=2}^{N-1}\left(E\pi_{\alpha_n}\big|S_{n-1} \sim S_n\right)P(S_{n-1} \sim S_n) + \left(E\pi_{\alpha_N}\big|S_{N-1}\right)P(S_{N-1}) + \sum_{n=1}^{N-1}\tau_n P(S_n)
\end{aligned} \quad (8)$$

as desired.

## 2 CASE $N = 2$ (TWO ALGORITHMS)

In this case

$$T_{\pi_{\alpha_1},\pi_{\alpha_2}}(\tau_1)(\omega) = \begin{cases} \pi_{\alpha_1}(\omega) & \pi_{\alpha_1}(\omega) \leq \tau_1 \\ \tau_1 + \pi_{\alpha_2}(\omega) & \tau_1 < \pi_{\alpha_1}(\omega) \end{cases} \quad (9)$$

$$ET_{\pi_{\alpha_1},\pi_{\alpha_2}}(\tau_1) = \left(E\pi_{\alpha_1}\big|\Omega \sim S_1\right)P(\Omega \sim S_1) + \left(E\pi_{\alpha_2}\big|S_1\right)P(S_1) + \tau_1 P(S_1) \quad (10)$$

and

$$S_1 = \{\omega \in \Omega; \tau_1 < \pi_{\alpha_1}(\omega)\}, \quad (11)$$

so

$$ET_{\pi_{\alpha_1},\pi_{\alpha_2}}(\tau_1) = \left(E\pi_{\alpha_1}\big|\pi_{\alpha_1} \leq \tau_1\right)P(\pi_{\alpha_1} \leq \tau_1) + \left(\left(E\pi_{\alpha_2}\big|\tau_1 < \pi_{\alpha_1}\right) + \tau_1\right)P(\tau_1 < \pi_{\alpha_1}) \quad (12)$$

### 2.1 $ET_{\pi_{\alpha_1},\pi_{\alpha_2}}(\tau_1)$ WHEN JOINT DENSITY IS $f_{\pi_{\alpha_1},\pi_{\alpha_2}}(x,y)$

In this case,

$$ET_{\pi_{\alpha_1},\pi_{\alpha_2}}(\tau_1) = \int_0^{\tau_1}\int_0^{\infty} x f_{\pi_{\alpha_1},\pi_{\alpha_2}}(x,y)\,dy\,dx + \int_{\tau_1}^{\infty}\int_0^{\infty}(\tau_1+y) f_{\pi_{\alpha_1},\pi_{\alpha_2}}(x,y)\,dy\,dx \quad (13)$$

## 2.11 EXAMPLE

Suppose the joint density of completion times for the two algorithms is given by

$$f_{\pi_{\alpha_1},\pi_{\alpha_2}}(x,y) = 12xy \exp\left(-(x+y)^2\right) \tag{14}$$

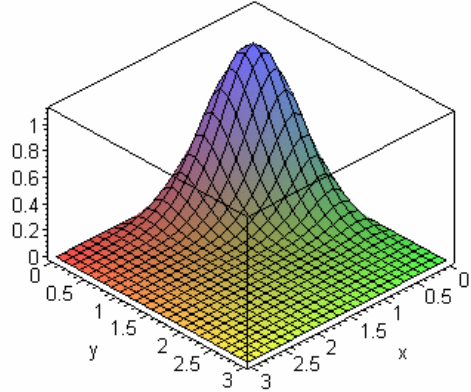

Figure 1. $f_{\pi_{\alpha_1},\pi_{\alpha_2}}$

Then

$$ET_{\pi_{\alpha_1},\pi_{\alpha_2}}(\tau_1) = \sqrt{\pi}\left(\text{erf}(\tau_1)-1\right)\left(\tau_1^4 + \tfrac{3}{2}\tau_1^2\right) + \left(\tau_1^3 + \tau_1\right)\exp\left(-\tau_1^2\right) + \tfrac{3}{8}\sqrt{\pi} \tag{15}$$

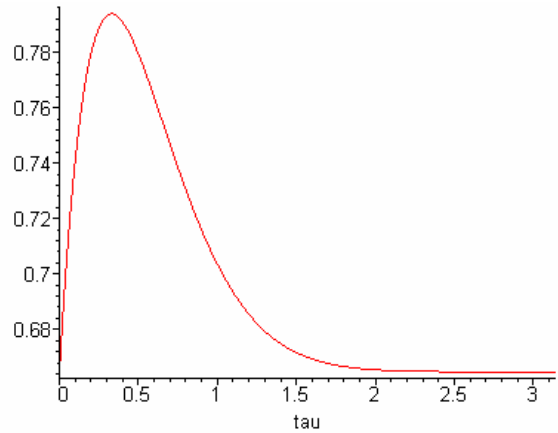

Figure 2. $ET_{\pi_{\alpha_1},\pi_{\alpha_2}}$

## 2.12 EXAMPLE

Suppose the joint density of completion times for the two algorithms is given by

$$f_{\pi_{\alpha_1},\pi_{\alpha_2}}(x,y) = 48xy\exp(-4x^2 - 3y^2) \tag{16}$$

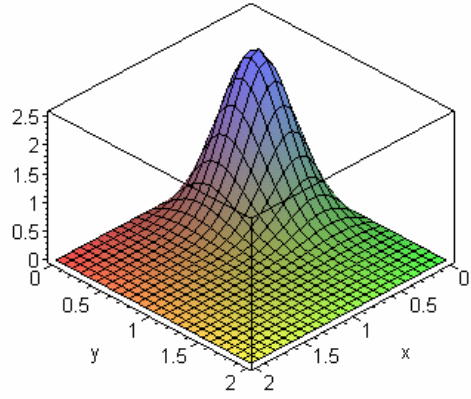

Figure 3. $f_{\pi_{\alpha_1},\pi_{\alpha_2}}$

Then

$$ET_{\pi_{\alpha_1},\pi_{\alpha_2}}(\tau_1) = \tfrac{1}{4}\sqrt{\pi}\,\mathrm{erf}(2\tau_1) + \tfrac{1}{6}\sqrt{3\pi}\exp(-4\tau_1^2) \tag{17}$$

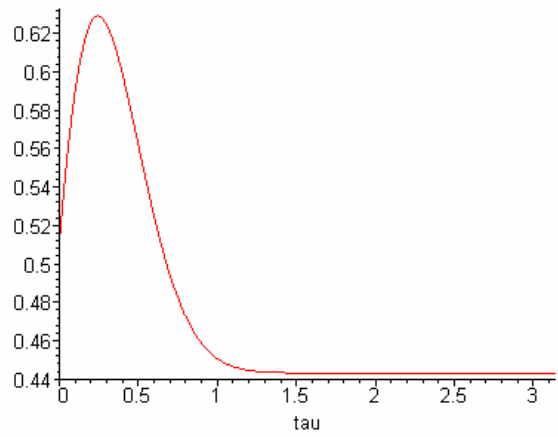

Figure 4. $ET_{\pi_{\alpha_1},\pi_{\alpha_2}}$

### 2.13 EXAMPLE

Suppose the joint density of completion times for the two algorithms is given by

$$f_{\pi_{\alpha_1},\pi_{\alpha_2}}(x,y) = .02217911969436783 0844\, xy \left( \begin{array}{l} \exp(-(x-1)^2 - (y-7)^2) \\ +\exp(-(x-7)^2 - (y-1)^2) \end{array} \right) \tag{18}$$

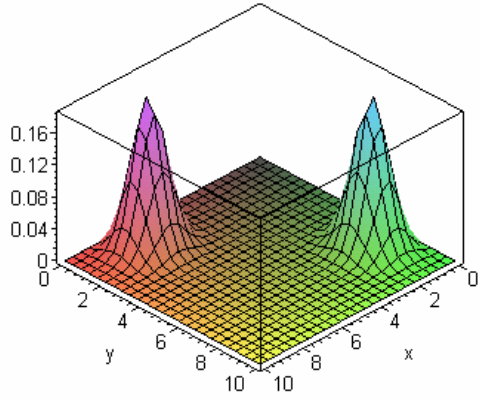

Figure 5. $f_{\pi_{\alpha_1},\pi_{\alpha_2}}$

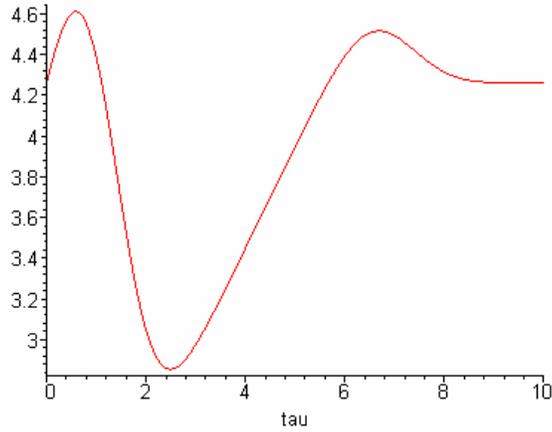

Figure 6. $ET_{\pi_{\alpha_1},\pi_{\alpha_2}}$

The minimum occurs at $\tau_1 \doteq 2.492$. Note that if $\tau_1 \doteq 2.492$, then $ET_{\pi_{\alpha_1},\pi_{\alpha_2}}(\tau_1) \doteq 2.854$, while $E\pi_{\alpha_1} \doteq 4.260$ and $E\pi_{\alpha_2} \doteq 4.260$. In this case the derived algorithm has better mean execution time than either of the original algorithms. Its mean execution time is approximately 33% less than that of either of the original algorithms.

<u>Notation</u>: In the following, wherever $B_1, B_2, \cdots, B_M$ is found in a context requiring a Boolean expression, it means the conjunction $B_1 \wedge B_2 \wedge \cdots \wedge B_M$ of the Boolean expressions $B_1, B_2, \cdots, B_M$.

Notation: In the following, if $B$ is a Boolean expression, then $(B) \equiv \begin{cases} 1 & B \text{ is true} \\ 0 & B \text{ is false} \end{cases}$.

In particular we define $(a \leq x < b) \equiv \begin{cases} 0 & x < a \\ 1 & a \leq x, x < b \\ 0 & b \leq x \end{cases}$.

**2.14** EXAMPLE

Suppose the joint density of completion times for the two algorithms is given by

$$f_{\pi_{\alpha_1}, \pi_{\alpha_2}}(x, y) = \tfrac{1}{12}(1 \leq x < 3)(4 \leq y < 7) \\ + \tfrac{1}{12}(5 \leq x < 8)(2 \leq y < 4)$$

(19)

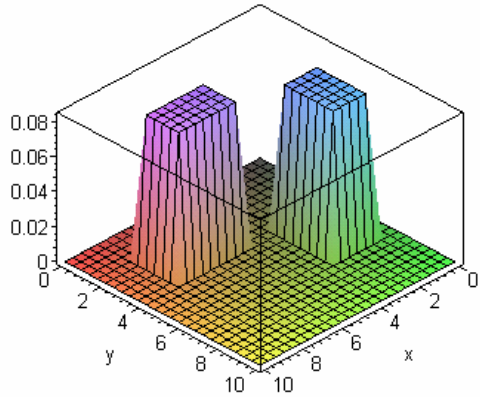

Figure 7. $f_{\pi_{\alpha_1}, \pi_{\alpha_2}}$

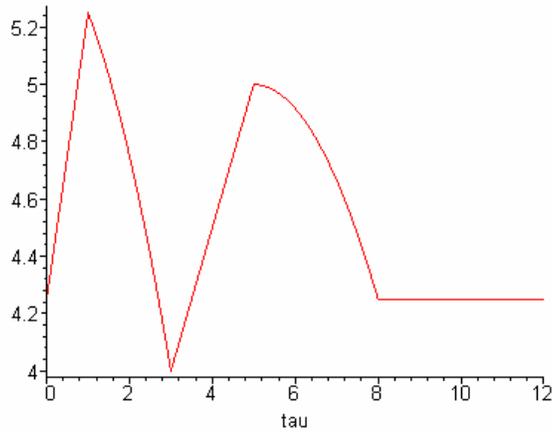

Figure 8. $ET_{\pi_{\alpha_1}, \pi_{\alpha_2}}$

The minimum occurs at $\tau_1 = 3$. Note that if $\tau_1 = 3$, then $ET_{\pi_{\alpha_1},\pi_{\alpha_2}}(\tau_1) = 4$, while $E\pi_{\alpha_1} = 4.25$ and $E\pi_{\alpha_2} = 4.25$. In this case the derived algorithm has better mean execution time than either of the original algorithms. Its mean execution time is approximately 6% less than that of either of the original algorithms.

**2.15** EXAMPLE

Suppose the joint density of completion times for the two algorithms is given by

$$f_{\pi_{\alpha_1},\pi_{\alpha_2}}(x,y) = \exp(-x-y) \tag{20}$$

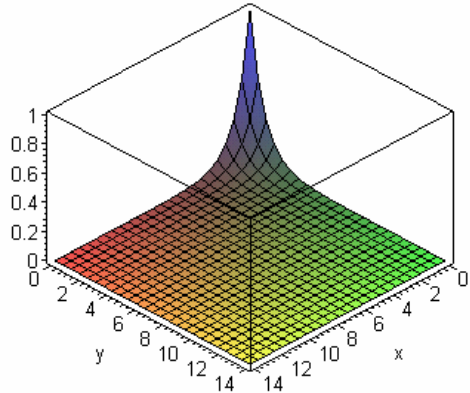

Figure 9. $f_{\pi_{\alpha_1},\pi_{\alpha_2}}$

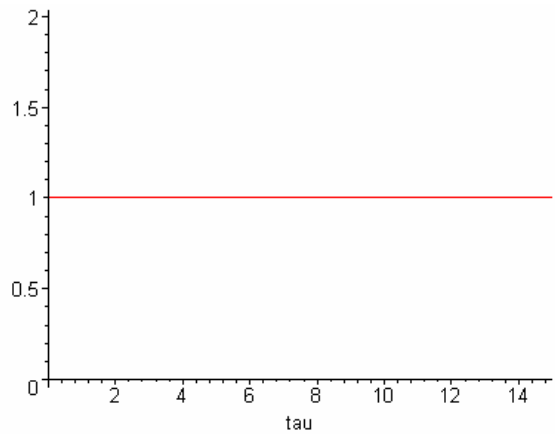

Figure 10. $ET_{\pi_{\alpha_1},\pi_{\alpha_2}}$

Note that for any choice of $\tau_1$, then $ET_{\pi_{\alpha_1},\pi_{\alpha_2}}(\tau_1)=1$, while $E\pi_{\alpha_1}=1$ and $E\pi_{\alpha_2}=1$. In this case the derived algorithm has exactly the same mean execution time as do the original algorithms, so a derived algorithm would be of no benefit.

**2.16 $ET_{\pi_{\alpha_1},\pi_{\alpha_2}}(\tau_1)$ DOES NOT ALWAYS EXIST**

If we take $f_{\pi_{\alpha_1},\pi_{\alpha_2}}(x,y) = \dfrac{1}{(x+1)^2(y+1)^2}$, then

$$\int_0^{\tau_1}\int_0^{\infty} x f_{\pi_{\alpha_1},\pi_{\alpha_2}}(x,y)\,dy\,dx + \int_{\tau_1}^{\infty}\int_0^{\infty}(\tau_1+y) f_{\pi_{\alpha_1},\pi_{\alpha_2}}(x,y)\,dy\,dx = \infty \qquad (21)$$

so in this case $ET_{\pi_{\alpha_1},\pi_{\alpha_2}}(\tau_1)$ does not exist.

**2.2 $ET_{\pi_{\alpha_1},\pi_{\alpha_2}}(\tau_1)$ WHEN JOINT DENSITY IS OF THE FORM**

$$f_{\pi_{\alpha_1},\pi_{\alpha_2}}(x,y) = f_{\pi_{\alpha_1}}(x) f_{\pi_{\alpha_2}}(y)$$

$$\begin{aligned}
ET_{\pi_{\alpha_1},\pi_{\alpha_2}}(\tau_1) &= \int_0^{\tau_1} x f_{\pi_{\alpha_1}}(x)\,dx + P(\tau_1 < \pi_{\alpha_1})(\tau_1 + E\pi_{\alpha_2}) \\
&= E(\pi_{\alpha_1}|\pi_{\alpha_1}\le\tau_1) P(\pi_{\alpha_1}\le\tau_1) + (\tau_1+E\pi_{\alpha_2}) P(\tau_1<\pi_{\alpha_1})
\end{aligned} \qquad (22)$$

**2.3 $ET_{\pi_{\alpha_1},\pi_{\alpha_2}}(\tau_1)$ WHEN JOINT DENSITY IS OF THE FORM**

$$f_{\pi_{\alpha_1},\pi_{\alpha_2}}(x,y) = \frac{1+ax+by+cx^2+dy^2}{1+a+b+2c+2d}\exp(-x-y)$$

If $0\le a,b,c,d$ then $f_{\pi_{\alpha_1},\pi_{\alpha_2}}(x,y) = \dfrac{1+ax+by+cx^2+dy^2}{1+a+b+2c+2d}\exp(-x-y)$ is a density function over $(x,y)\in[0,\infty)\times[0,\infty)$. Accordingly,

$$ET_{\pi_{\alpha_1},\pi_{\alpha_2}}(\tau_1) = \frac{1+2a+b+6c+2d-(2c\tau_1+a-b+4c-4d)\exp(-\tau_1)}{1+a+b+2c+2d} \qquad (23)$$

## 2.4 $ET_{\pi_{\alpha_1},\pi_{\alpha_2}}(\tau_1)$ WHEN JOINT DENSITY IS OF THE FORM

$$f_{\pi_{\alpha_1},\pi_{\alpha_2}}(x,y) = \frac{2c}{(1+x)^2(1+y)^3} + \sum_{m=0}^{\infty}\sum_{n=0}^{\infty} d(m,n)\frac{(m+2)(n+2)}{(1+x)^{m+3}(1+y)^{n+3}}$$

If $0 \leq d(m,n)$, $0 \leq c$, $c + \sum_{m=0}^{\infty}\sum_{n=0}^{\infty} d(m,n) = 1$, then

$$f_{\pi_{\alpha_1},\pi_{\alpha_2}}(x,y) = \frac{2c}{(1+x)^2(1+y)^3} + \sum_{m=0}^{\infty}\sum_{n=0}^{\infty} d(m,n)\frac{(m+2)(n+2)}{(1+x)^{m+3}(1+y)^{n+3}} \tag{24}$$

is a density function over $(x,y) \in [0,\infty) \times [0,\infty)$. A straightforward calculation yields

$$\begin{aligned}ET(\tau_1) &= \frac{1}{2}\sum_{n=0}^{\infty} d(1,n) + \sum_{m=2}^{\infty}\frac{1}{m+1}\sum_{n=0}^{\infty} d(m,n) + \frac{c}{1+\tau_1} + \frac{\tau_1}{1+\tau_1}\sum_{n=0}^{\infty} d(0,n) \\ &+ \sum_{m=2}^{\infty}\frac{1}{(1+\tau_1)^m}\sum_{n=0}^{\infty}\left(\frac{1}{n+1}d(m-2,n) - \frac{1}{m}d(m-1,n)\right) + c\ln(1+\tau_1)\end{aligned} \tag{25}$$

## 2.5 $ET_{\pi_{\alpha_1},\pi_{\alpha_2}}(\tau_1)$ WHEN JOINT DENSITY IS OF THE FORM

$$f_{\pi_{\alpha_1},\pi_{\alpha_2}}(x,y) = \sum_{n=1}^{N} k_n (a_n \leq x < b_n)(c_n \leq y < d_n)$$

**Theorem 2:** If $\sum_{n=1}^{N} k_n (b_n - a_n)(d_n - c_n) = 1$ and $0 < k_n$ for $n = 1, 2, \cdots, N$, and

$$f_{\pi_{\alpha_1},\pi_{\alpha_2}}(x,y) = \sum_{n=1}^{N} k_n (a_n \leq x < b_n)(c_n \leq y < d_n), \text{ then}$$

$$ET_{\pi_{\alpha_1},\pi_{\alpha_2}}(\tau_1) = \tau_1^2 \left( -\tfrac{1}{2} \sum_{\substack{n=1 \\ a_n < \tau_1 < b_n}}^{N} k_n (d_n - c_n) \right)$$

$$+\tau_1 \left( \begin{array}{l} \sum_{\substack{n=1 \\ \tau_1 \le a_n}}^{N} k_n (b_n - a_n)(d_n - c_n) \\ + \sum_{\substack{n=1 \\ a_n < \tau_1 < b_n}}^{N} k_n \left(b_n - \tfrac{1}{2}(d_n + c_n)\right)(d_n - c_n) \end{array} \right)$$

$$+ \tfrac{1}{2} \sum_{\substack{n=1 \\ \tau_1 \le a_n}}^{N} k_n (b_n - a_n)(d_n^2 - c_n^2) \tag{26}$$

$$+ \tfrac{1}{2} \sum_{\substack{n=1 \\ a_n < \tau_1 < b_n}}^{N} k_n \left(b_n(d_n + c_n) - a_n^2\right)(d_n - c_n)$$

$$+ \tfrac{1}{2} \sum_{\substack{n=1 \\ b_n \le \tau_1}}^{N} k_n (b_n^2 - a_n^2)(d_n - c_n)$$

<u>Notation</u>: In the following, if $B(n)$ is a Boolean expression, then $\sum_{\substack{n=1 \\ B(n)}}^{N} s_n \equiv \sum_{n=1}^{N} (B(n)) s_n$.

<u>Proof</u>: If $\sum_{n=1}^{N} k_n (b_n - a_n)(d_n - c_n) = 1$ and $0 < k_n$ for $n = 1, 2, \cdots, N$, then

$f_{\pi_{\alpha_1},\pi_{\alpha_2}}(x, y) = \sum_{n=1}^{N} k_n (a_n \le x < b_n)(c_n \le y < d_n)$ is a density function over

$(x, y) \in [0, \infty) \times [0, \infty)$. Now note that

$$ET_{\pi_{\alpha_1},\pi_{\alpha_2}}(\tau_1) = \int_0^{\tau_1} \int_0^{\infty} x f_{\pi_{\alpha_1},\pi_{\alpha_2}}(x, y) \, dy \, dx + \int_{\tau_1}^{\infty} \int_0^{\infty} (\tau_1 + y) f_{\pi_{\alpha_1},\pi_{\alpha_2}}(x, y) \, dy \, dx$$

$$= \int_0^{\tau_1} \int_0^{\infty} x \sum_{n=1}^{N} k_n (a_n \le x < b_n)(c_n \le y < d_n) \, dy \, dx$$

$$+ \int_{\tau_1}^{\infty} \int_0^{\infty} (\tau_1 + y) \sum_{n=1}^{N} k_n (a_n \le x < b_n)(c_n \le y < d_n) \, dy \, dx \tag{27}$$

$$= \sum_{n=1}^{N} \int_{0}^{\tau_1} \int_{0}^{\infty} k_n x \left(a_n \leq x < b_n\right)\left(c_n \leq y < d_n\right) dy dx$$

$$+ \sum_{n=1}^{N} \int_{\tau_1}^{\infty} \int_{0}^{\infty} k_n \left(\tau_1 + y\right)\left(a_n \leq x < b_n\right)\left(c_n \leq y < d_n\right) dy dx \qquad (28)$$

$$= \sum_{n=1}^{N} k_n \left(d_n - c_n\right) \int_{0}^{\tau_1} x \left(a_n \leq x < b_n\right) dx$$

$$+ \sum_{n=1}^{N} k_n \int_{\tau_1}^{\infty} \left(a_n \leq x < b_n\right) dx \int_{c_n}^{d_n} \left(\tau_1 + y\right) dy$$

$$= \sum_{n=1}^{N} k_n \left(d_n - c_n\right) \int_{0}^{\tau_1} x \left(a_n \leq x < b_n\right) dx$$

$$+ \sum_{n=1}^{N} k_n \int_{\tau_1}^{\infty} \left(a_n \leq x < b_n\right) dx \left(\frac{\left(\tau_1 + d_n\right)^2}{2} - \frac{\left(\tau_1 + c_n\right)^2}{2}\right)$$

$$= \sum_{\substack{n=1 \\ \tau_1 \leq a_n}}^{N} k_n \left(d_n - c_n\right) \int_{0}^{\tau_1} x \left(a_n \leq x < b_n\right) dx$$

$$+ \sum_{\substack{n=1 \\ a_n < \tau_1 < b_n}}^{N} k_n \left(d_n - c_n\right) \int_{0}^{\tau_1} x \left(a_n \leq x < b_n\right) dx$$

$$+ \sum_{\substack{n=1 \\ b_n \leq \tau_1}}^{N} k_n \left(d_n - c_n\right) \int_{0}^{\tau_1} x \left(a_n \leq x < b_n\right) dx$$

$$+ \frac{1}{2} \sum_{\substack{n=1 \\ \tau_1 \leq a_n}}^{N} k_n \left(d_n - c_n\right)\left(2\tau_1 + c_n + d_n\right) \int_{\tau_1}^{\infty} \left(a_n \leq x < b_n\right) dx$$

$$+ \frac{1}{2} \sum_{\substack{n=1 \\ a_n < \tau_1 < b_n}}^{N} k_n \left(d_n - c_n\right)\left(2\tau_1 + c_n + d_n\right) \int_{\tau_1}^{\infty} \left(a_n \leq x < b_n\right) dx \qquad (29)$$

$$+ \frac{1}{2} \sum_{\substack{n=1 \\ b_n \leq \tau_1}}^{N} k_n \left(d_n - c_n\right)\left(2\tau_1 + c_n + d_n\right) \int_{\tau_1}^{\infty} \left(a_n \leq x < b_n\right) dx$$

$$ET_{\pi_{\alpha_1},\pi_{\alpha_2}}(\tau_1) = \sum_{\substack{n=1 \\ a_n < \tau_1 < b_n}}^{N} k_n(d_n - c_n)\int_0^{\tau_1} x(a_n \leq x < b_n)dx$$

$$+ \sum_{\substack{n=1 \\ b_n \leq \tau_1}}^{N} k_n(d_n - c_n)\int_0^{\tau_1} x(a_n \leq x < b_n)dx$$

$$+ \tfrac{1}{2}\sum_{\substack{n=1 \\ \tau_1 \leq a_n}}^{N} k_n(d_n - c_n)(2\tau_1 + c_n + d_n)\int_{\tau_1}^{\infty}(a_n \leq x < b_n)dx$$

$$+ \tfrac{1}{2}\sum_{\substack{n=1 \\ a_n < \tau_1 < b_n}}^{N} k_n(d_n - c_n)(2\tau_1 + c_n + d_n)\int_{\tau_1}^{\infty}(a_n \leq x < b_n)dx$$

$$= \sum_{\substack{n=1 \\ a_n < \tau_1 < b_n}}^{N} k_n(d_n - c_n)\int_{a_n}^{\tau_1} x\,dx + \sum_{\substack{n=1 \\ b_n \leq \tau_1}}^{N} k_n(d_n - c_n)\int_{a_n}^{b_n} x\,dx$$

$$+ \tfrac{1}{2}\sum_{\substack{n=1 \\ \tau_1 \leq a_n}}^{N} k_n(d_n - c_n)(2\tau_1 + c_n + d_n)\int_{a_n}^{b_n} 1\,dx$$

$$+ \tfrac{1}{2}\sum_{\substack{n=1 \\ a_n < \tau_1 < b_n}}^{N} k_n(d_n - c_n)(2\tau_1 + c_n + d_n)\int_{\tau_1}^{b_n} 1\,dx$$

$$= \sum_{\substack{n=1 \\ a_n < \tau_1 < b_n}}^{N} k_n(d_n - c_n)\left(\tfrac{1}{2}\tau_1^2 - \tfrac{1}{2}a_n^2\right) + \sum_{\substack{n=1 \\ b_n \leq \tau_1}}^{N} k_n(d_n - c_n)\tfrac{1}{2}(b_n^2 - a_n^2) \quad (30)$$

$$+ \tfrac{1}{2}\sum_{\substack{n=1 \\ \tau_1 \leq a_n}}^{N} k_n(d_n - c_n)(2\tau_1 + c_n + d_n)(b_n - a_n)$$

$$+ \tfrac{1}{2}\sum_{\substack{n=1 \\ a_n < \tau_1 < b_n}}^{N} k_n(d_n - c_n)(2\tau_1 + c_n + d_n)(b_n - \tau_1)$$

$$= \tau_1^2 \left( -\tfrac{1}{2} \sum_{\substack{n=1 \\ a_n < \tau_1 < b_n}}^{N} k_n (d_n - c_n) \right)$$

$$+ \tau_1 \left( \begin{array}{l} \displaystyle\sum_{\substack{n=1 \\ \tau_1 \le a_n}}^{N} k_n (b_n - a_n)(d_n - c_n) \\ + \displaystyle\sum_{\substack{n=1 \\ a_n < \tau_1 < b_n}}^{N} k_n \left(b_n - \tfrac{1}{2}(d_n + c_n)\right)(d_n - c_n) \end{array} \right)$$

$$+ \tfrac{1}{2} \sum_{\substack{n=1 \\ \tau_1 \le a_n}}^{N} k_n (b_n - a_n)(d_n^2 - c_n^2)$$

$$+ \tfrac{1}{2} \sum_{\substack{n=1 \\ a_n < \tau_1 < b_n}}^{N} k_n \left(b_n (d_n + c_n) - a_n^2\right)(d_n - c_n)$$

$$+ \tfrac{1}{2} \sum_{\substack{n=1 \\ b_n \le \tau_1}}^{N} k_n (b_n^2 - a_n^2)(d_n - c_n) \tag{31}$$

$$ET_{\pi_{\alpha_1}, \pi_{\alpha_2}}(\tau_1) = \tau_1^2 \left( -\tfrac{1}{2} \sum_{\substack{n=1 \\ a_n < \tau_1 < b_n}}^{N} k_n (d_n - c_n) \right)$$

$$+ \tau_1 \left( \begin{array}{l} \displaystyle\sum_{\substack{n=1 \\ \tau_1 \le a_n}}^{N} k_n (b_n - a_n)(d_n - c_n) \\ + \displaystyle\sum_{\substack{n=1 \\ a_n < \tau_1 < b_n}}^{N} k_n \left(b_n - \tfrac{1}{2}(d_n + c_n)\right)(d_n - c_n) \end{array} \right)$$

$$+ \tfrac{1}{2} \sum_{\substack{n=1 \\ \tau_1 \le a_n}}^{N} k_n (b_n - a_n)(d_n^2 - c_n^2)$$

$$+ \tfrac{1}{2} \sum_{\substack{n=1 \\ a_n < \tau_1 < b_n}}^{N} k_n \left(b_n (d_n + c_n) - a_n^2\right)(d_n - c_n)$$

$$+ \tfrac{1}{2} \sum_{\substack{n=1 \\ b_n \le \tau_1}}^{N} k_n (b_n^2 - a_n^2)(d_n - c_n) \tag{32}$$

In particular,

$$ET_{\pi_{\alpha_1},\pi_{\alpha_2}}(a_i) = a_i^2 \left( -\tfrac{1}{2} \sum_{\substack{n=1 \\ a_n < a_i < b_n}}^{N} k_n (d_n - c_n) \right)$$

$$+ a_i \left( \begin{array}{l} \sum_{\substack{n=1 \\ a_i \le a_n}}^{N} k_n (b_n - a_n)(d_n - c_n) \\ + \sum_{\substack{n=1 \\ a_n < a_i < b_n}}^{N} k_n \left(b_n - \tfrac{1}{2}(d_n + c_n)\right)(d_n - c_n) \end{array} \right)$$

$$+ \tfrac{1}{2} \sum_{\substack{n=1 \\ a_i \le a_n}}^{N} k_n (b_n - a_n)(d_n^2 - c_n^2) \qquad (33)$$

$$+ \tfrac{1}{2} \sum_{\substack{n=1 \\ a_n < a_i < b_n}}^{N} k_n \left(b_n (d_n + c_n) - a_n^2\right)(d_n - c_n)$$

$$+ \tfrac{1}{2} \sum_{\substack{n=1 \\ b_n \le a_i}}^{N} k_n (b_n^2 - a_n^2)(d_n - c_n)$$

$$ET_{\pi_{\alpha_1},\pi_{\alpha_2}}(b_i) = b_i^2 \left( -\tfrac{1}{2} \sum_{\substack{n=1 \\ a_n < b_i < b_n}}^{N} k_n (d_n - c_n) \right)$$

$$+ b_i \left( \begin{array}{l} \sum_{\substack{n=1 \\ b_i \le a_n}}^{N} k_n (b_n - a_n)(d_n - c_n) \\ + \sum_{\substack{n=1 \\ a_n < b_i < b_n}}^{N} k_n \left(b_n - \tfrac{1}{2}(d_n + c_n)\right)(d_n - c_n) \end{array} \right)$$

$$+ \tfrac{1}{2} \sum_{\substack{n=1 \\ b_i \le a_n}}^{N} k_n (b_n - a_n)(d_n^2 - c_n^2) \qquad (34)$$

$$+ \tfrac{1}{2} \sum_{\substack{n=1 \\ a_n < b_i < b_n}}^{N} k_n \left(b_n (d_n + c_n) - a_n^2\right)(d_n - c_n)$$

$$+ \tfrac{1}{2} \sum_{\substack{n=1 \\ b_n \le b_i}}^{N} k_n (b_n^2 - a_n^2)(d_n - c_n)$$

It is straightforward to show that $ET_{\pi_{\alpha_1},\pi_{\alpha_2}}(\tau_1)$ attains a global minimum at one of the points $\{a_1, a_2, \cdots a_N, b_1, b_2, \cdots b_N\}$. Indeed, $ET_{\pi_{\alpha_1},\pi_{\alpha_2}}(\tau_1)$ is a continuous, piecewise quadratic function. Notice that the set of points of connection of the pieces is a subset of $\{a_1, a_2, \cdots a_N, b_1, b_2, \cdots b_N\}$. Each piece is either linear or is quadratic with a negative second derivative. We can thus replace each quadratic piece with a linear piece connecting the endpoints of the quadratic piece, without altering the global minimum of $ET_{\pi_{\alpha_1}, \pi_{\alpha_2}}(\tau_1)$. After replacing each quadratic piece with the appropriate linear piece, we then have a continuous piecewise linear function whose global minimum is the same as that of $ET_{\pi_{\alpha_1}, \pi_{\alpha_2}}(\tau_1)$. But of course the global minimum of a continuous piecewise linear function is attained at one of its vertices. These vertices are a subset of the set of points of connection $\{a_1, a_2, \cdots a_N, b_1, b_2, \cdots b_N\}$, as desired.

This global minimum is given by

$$\min \left\{ \begin{array}{l} ET_{\pi_{\alpha_1},\pi_{\alpha_2}}(a_1), ET_{\pi_{\alpha_1},\pi_{\alpha_2}}(a_2), \cdots, ET_{\pi_{\alpha_1},\pi_{\alpha_2}}(a_N) \\ , ET_{\pi_{\alpha_1},\pi_{\alpha_2}}(b_1), ET_{\pi_{\alpha_1},\pi_{\alpha_2}}(b_2), \cdots, ET_{\pi_{\alpha_1},\pi_{\alpha_2}}(b_N) \end{array} \right\} \tag{35}$$

This minimum can be computed in $O(N^2)$ time.

MAXIMUM LIKELIHOOD ESTIMATION

Suppose $f_{\pi_{\alpha_1},\pi_{\alpha_2}}(x,y) = \sum_{n=1}^{N} k_n (a_n \leq x < b_n)(c_n \leq y < d_n)$

with the following conditions:

1. $k_n > 0$ for $1 \leq n \leq N$,

2. $a_n < b_n$ for $1 \leq n \leq N$,

3. $c_n < d_n$ for $1 \leq n \leq N$,

4. The boxes $B_n \equiv [a_n, b_n) \times [c_n, d_n)$ for $1 \leq n \leq N$ are disjoint,

5. $\sum_{n=1}^{N} k_n (b_n - a_n)(d_n - c_n) = 1$.

Then $f_{\pi_{\alpha_1}, \pi_{\alpha_2}}$ is a joint density function.

Suppose next that we have observed the performance of two equivalent algorithms $\alpha$ and $\beta$ over a (finite) sample set $\Omega_s \subset \Omega$. That is, for each task $\omega \in \Omega_s$ we have observed the values $\pi_\alpha(\omega)$ and $\pi_\beta(\omega)$ representing the time that algorithms $\alpha$ and $\beta$ actually took to process the task $\omega$. We now present a maximum-likelihood procedure to find the "best fitting" joint density function of the form

$$f_{\pi_{\alpha_1}, \pi_{\alpha_2}}(x, y) = \sum_{n=1}^{N} k_n (a_n \leq x < b_n)(c_n \leq y < d_n),$$

subject to the five conditions above. Let $\{(x_1, y_1), (x_2, y_2), \cdots, (x_P, y_P)\}$ be the data observed, where $x_j$ and $y_j$ are the durations required by algorithms $\alpha$ and $\beta$ respectively, to process $\omega_j \in \Omega_s$, for $1 \leq j \leq P$, with $P = |\Omega_s|$. Our performance function is defined as

$$g(k_1, k_2, \cdots, k_N) \equiv \prod_{m=1}^{P} f_{\pi_{\alpha_1}, \pi_{\alpha_2}}(x_m, y_m) = \prod_{m=1}^{P} \sum_{n=1}^{N} k_n (a_n \leq x_m < b_n)(c_n \leq y_m < d_n) = k_1^{S_1} k_2^{S_2} \cdots k_N^{S_N}$$

where

$$S_j \equiv \left| \{(x, y) \in \{(x_1, y_1), (x_2, y_2), \cdots, (x_P, y_P)\}; a_j \leq x < b_j, c_j \leq y < d_j \} \right|$$

We form as usual the Lagrange multiplier equations

$\lambda S_j \dfrac{1}{k_j} + (b_j - a_j)(d_j - c_j) = 0$ for $1 \leq j \leq P$. We have immediately that

$$k_j = -\frac{\lambda S_j}{(b_j - a_j)(d_j - c_j)}$$

and recalling the constraint

$$1 = \sum_{n=1}^{N} k_n (b_n - a_n)(d_n - c_n)$$

we infer

$$1 = \sum_{n=1}^{N} -\lambda S_n = -\lambda P$$

whence

$$\lambda = -\frac{1}{P}$$

thus

$$k_j = \frac{S_j}{P(b_j - a_j)(d_j - c_j)}$$

Substituting into (32), we get

$$ET_{\pi_{\alpha_1},\pi_{\alpha_2}}(\tau_1) = \tau_1^2 \left( -\frac{1}{2} \sum_{\substack{n=1 \\ a_n < \tau_1 < b_n}}^{N} \frac{S_n}{P(b_n - a_n)(d_n - c_n)} (d_n - c_n) \right)$$

$$+ \tau_1 \left( \sum_{\substack{n=1 \\ \tau_1 \leq a_n}}^{N} \frac{S_n}{P(b_n - a_n)(d_n - c_n)} (b_n - a_n)(d_n - c_n) \right.$$

$$\left. + \sum_{\substack{n=1 \\ a_n < \tau_1 < b_n}}^{N} \frac{S_n}{P(b_n - a_n)(d_n - c_n)} \left(b_n - \tfrac{1}{2}(d_n + c_n)\right)(d_n - c_n) \right)$$

$$+ \tfrac{1}{2} \sum_{\substack{n=1 \\ \tau_1 \leq a_n}}^{N} \frac{S_n}{P(b_n - a_n)(d_n - c_n)} (b_n - a_n)(d_n^2 - c_n^2)$$

$$+ \tfrac{1}{2} \sum_{\substack{n=1 \\ a_n < \tau_1 < b_n}}^{N} \frac{S_n}{P(b_n - a_n)(d_n - c_n)} \left(b_n(d_n + c_n) - a_n^2\right)(d_n - c_n)$$

$$+ \tfrac{1}{2} \sum_{\substack{n=1 \\ b_n \leq \tau_1}}^{N} \frac{S_n}{P(b_n - a_n)(d_n - c_n)} (b_n^2 - a_n^2)(d_n - c_n)$$

$$= \tau_1^2 \left( -\tfrac{1}{2} \sum_{\substack{n=1 \\ a_n < \tau_1 < b_n}}^{N} \frac{S_n}{P(b_n - a_n)} \right) + \tau_1 \left( \sum_{\substack{n=1 \\ \tau_1 \leq a_n}}^{N} \frac{S_n}{P} + \sum_{\substack{n=1 \\ a_n < \tau_1 < b_n}}^{N} \frac{S_n}{P(b_n - a_n)} (b_n - \tfrac{1}{2}(d_n + c_n)) \right)$$

$$+ \tfrac{1}{2} \sum_{\substack{n=1 \\ \tau_1 \leq a_n}}^{N} \frac{S_n}{P}(d_n + c_n) + \tfrac{1}{2} \sum_{\substack{n=1 \\ a_n < \tau_1 < b_n}}^{N} \frac{S_n}{P(b_n - a_n)} (b_n(d_n + c_n) - a_n^2) + \tfrac{1}{2} \sum_{\substack{n=1 \\ b_n \leq \tau_1}}^{N} \frac{S_n}{P}(b_n + a_n)$$

$$ET_{\pi_{\alpha_1}, \pi_{\alpha_2}}(\tau_1) = -\tfrac{1}{2P} \tau_1^2 \sum_{\substack{n=1 \\ a_n < \tau_1 < b_n}}^{N} \frac{S_n}{b_n - a_n} + \tau_1 \frac{1}{P} \left( \sum_{\substack{n=1 \\ \tau_1 \leq a_n}}^{N} S_n + \sum_{\substack{n=1 \\ a_n < \tau_1 < b_n}}^{N} S_n \frac{b_n - \tfrac{1}{2}(d_n + c_n)}{b_n - a_n} \right)$$

$$+ \tfrac{1}{2P} \sum_{\substack{n=1 \\ a_n < \tau_1 < b_n}}^{N} \frac{S_n}{b_n - a_n} (b_n(d_n + c_n) - a_n^2) + \tfrac{1}{2P} \sum_{\substack{n=1 \\ \tau_1 \leq a_n}}^{N} S_n(d_n + c_n) + \tfrac{1}{2P} \sum_{\substack{n=1 \\ b_n \leq \tau_1}}^{N} S_n(b_n + a_n)$$

## 3 CONCLUSIONS

In this paper, we asked the following questions: Given two or more equivalent algorithms, is it ever possible to create a new derived algorithm whose mean execution time is less than that of all of the original algorithms? If so, how can such an algorithm be derived?

By giving examples in Section 2, we have shown that the answer to the first question is "yes." In Section 1, we gave an explicit construction of the derived algorithm.